\PassOptionsToPackage{unicode}{hyperref}
\PassOptionsToPackage{hyphens}{url}
\PassOptionsToPackage{dvipsnames,svgnames,x11names}{xcolor}
\documentclass[
]{article}
\usepackage{amsmath,amssymb}
\usepackage{lmodern}
\usepackage{iftex}
\ifPDFTeX
  \usepackage[T1]{fontenc}
  \usepackage[utf8]{inputenc}
  \usepackage{textcomp} 
\else 
  \usepackage{unicode-math}
  \defaultfontfeatures{Scale=MatchLowercase}
  \defaultfontfeatures[\rmfamily]{Ligatures=TeX,Scale=1}
\fi
\IfFileExists{upquote.sty}{\usepackage{upquote}}{}
\IfFileExists{microtype.sty}{
  \usepackage[]{microtype}
  \UseMicrotypeSet[protrusion]{basicmath} 
}{}
\makeatletter
\@ifundefined{KOMAClassName}{
  \IfFileExists{parskip.sty}{%
    \usepackage{parskip}
  }{
    \setlength{\parindent}{0pt}
    \setlength{\parskip}{6pt plus 2pt minus 1pt}}
}{
  \KOMAoptions{parskip=half}}
\makeatother
\usepackage{xcolor}
\usepackage{graphicx}
\makeatletter
\def\maxwidth{\ifdim\Gin@nat@width>\linewidth\linewidth\else\Gin@nat@width\fi}
\def\maxheight{\ifdim\Gin@nat@height>\textheight\textheight\else\Gin@nat@height\fi}
\makeatother
\setkeys{Gin}{width=\maxwidth,height=\maxheight,keepaspectratio}
\makeatletter
\def\fps@figure{htbp}
\makeatother
\providecommand{\tightlist}{%
  \setlength{\itemsep}{0pt}\setlength{\parskip}{0pt}}
\NewDocumentCommand\citeproctext{}{}
\NewDocumentCommand\citeproc{mm}{%
  \begingroup\def\citeproctext{#2}\cite{#1}\endgroup}
\makeatletter
 \let\@cite@ofmt\@firstofone
 \def\@biblabel#1{}
 \def\@cite#1#2{{#1\if@tempswa , #2\fi}}
\makeatother
\newlength{\cslhangindent}
\newlength{\csllabelwidth}
\newenvironment{CSLReferences}[2] 
 {\begin{list}{}{%
  \setlength{\itemindent}{0pt}
  \setlength{\leftmargin}{0pt}
  \setlength{\parsep}{0pt}
  \ifodd #1
   \setlength{\leftmargin}{\cslhangindent}
   \setlength{\itemindent}{-1\cslhangindent}
  \fi
  \setlength{\itemsep}{#2\baselineskip}}}
 {\end{list}}
\usepackage{calc}

\ifLuaTeX
\usepackage[bidi=basic]{babel}
\else
\usepackage[bidi=default]{babel}
\fi
\babelprovide[main,import]{american}

\def\languageshorthands#1{}
\ifLuaTeX
  \usepackage{selnolig}  
\fi
\IfFileExists{bookmark.sty}{\usepackage{bookmark}}{\usepackage{hyperref}}
\IfFileExists{xurl.sty}{\usepackage{xurl}}{} 
\hypersetup{
  pdftitle={Whole slide and microscopy image analysis with QuPath and
OMERO},
  pdfauthor={Léo Leplat, Alan O'Callaghan, Peter Bankhead},
  pdflang={en-US},
  colorlinks=true,
  linkcolor={Maroon},
  filecolor={Maroon},
  citecolor={Blue},
  urlcolor={Blue},
  pdfcreator={LaTeX via pandoc}}

\title{Whole slide and microscopy image analysis with QuPath and OMERO}

\definecolor{c53baa1}{RGB}{83,186,161}
\definecolor{c202826}{RGB}{32,40,38}
\def \rorglobalscale {0.1}
\newcommand{\rorlogo}{%
\begin{tikzpicture}[y=1cm, x=1cm, yscale=\rorglobalscale,xscale=\rorglobalscale, every node/.append style={scale=\rorglobalscale}, inner sep=0pt, outer sep=0pt]
  \begin{scope}[even odd rule,line join=round,miter limit=2.0,shift={(-0.025, 0.0216)}]
    \path[fill=c53baa1,nonzero rule,line join=round,miter limit=2.0] (1.8164, 3.012) -- (1.4954, 2.5204) -- (1.1742, 3.012) -- (1.8164, 3.012) -- cycle;
    \path[fill=c53baa1,nonzero rule,line join=round,miter limit=2.0] (3.1594, 3.012) -- (2.8385, 2.5204) -- (2.5172, 3.012) -- (3.1594, 3.012) -- cycle;
    \path[fill=c53baa1,nonzero rule,line join=round,miter limit=2.0] (1.1742, 0.0669) -- (1.4954, 0.5588) -- (1.8164, 0.0669) -- (1.1742, 0.0669) -- cycle;
    \path[fill=c53baa1,nonzero rule,line join=round,miter limit=2.0] (2.5172, 0.0669) -- (2.8385, 0.5588) -- (3.1594, 0.0669) -- (2.5172, 0.0669) -- cycle;
    \path[fill=c202826,nonzero rule,line join=round,miter limit=2.0] (3.8505, 1.4364).. controls (3.9643, 1.4576) and (4.0508, 1.5081) .. (4.1098, 1.5878).. controls (4.169, 1.6674) and (4.1984, 1.7642) .. (4.1984, 1.8777).. controls (4.1984, 1.9719) and (4.182, 2.0503) .. (4.1495, 2.1132).. controls (4.1169, 2.1762) and (4.0727, 2.2262) .. (4.0174, 2.2635).. controls (3.9621, 2.3006) and (3.8976, 2.3273) .. (3.824, 2.3432).. controls (3.7505, 2.359) and (3.6727, 2.367) .. (3.5909, 2.367) -- (2.9676, 2.367) -- (2.9676, 1.8688).. controls (2.9625, 1.8833) and (2.9572, 1.8976) .. (2.9514, 1.9119).. controls (2.9083, 2.0164) and (2.848, 2.1056) .. (2.7705, 2.1791).. controls (2.6929, 2.2527) and (2.6014, 2.3093) .. (2.495, 2.3487).. controls (2.3889, 2.3881) and (2.2728, 2.408) .. (2.1468, 2.408).. controls (2.0209, 2.408) and (1.905, 2.3881) .. (1.7986, 2.3487).. controls (1.6925, 2.3093) and (1.6007, 2.2527) .. (1.5232, 2.1791).. controls (1.4539, 2.1132) and (1.3983, 2.0346) .. (1.3565, 1.9436).. controls (1.3504, 2.009) and (1.3351, 2.0656) .. (1.3105, 2.1132).. controls (1.2779, 2.1762) and (1.2338, 2.2262) .. (1.1785, 2.2635).. controls (1.1232, 2.3006) and (1.0586, 2.3273) .. (0.985, 2.3432).. controls (0.9115, 2.359) and (0.8337, 2.367) .. (0.7519, 2.367) -- (0.1289, 2.367) -- (0.1289, 0.7562) -- (0.4837, 0.7562) -- (0.4837, 1.4002) -- (0.6588, 1.4002) -- (0.9956, 0.7562) -- (1.4211, 0.7562) -- (1.0118, 1.4364).. controls (1.1255, 1.4576) and (1.2121, 1.5081) .. (1.2711, 1.5878).. controls (1.2737, 1.5915) and (1.2761, 1.5954) .. (1.2787, 1.5991).. controls (1.2782, 1.5867) and (1.2779, 1.5743) .. (1.2779, 1.5616).. controls (1.2779, 1.4327) and (1.2996, 1.3158) .. (1.3428, 1.2113).. controls (1.3859, 1.1068) and (1.4462, 1.0176) .. (1.5237, 0.944).. controls (1.601, 0.8705) and (1.6928, 0.8139) .. (1.7992, 0.7744).. controls (1.9053, 0.735) and (2.0214, 0.7152) .. (2.1474, 0.7152).. controls (2.2733, 0.7152) and (2.3892, 0.735) .. (2.4956, 0.7744).. controls (2.6016, 0.8139) and (2.6935, 0.8705) .. (2.771, 0.944).. controls (2.8482, 1.0176) and (2.9086, 1.1068) .. (2.952, 1.2113).. controls (2.9578, 1.2253) and (2.9631, 1.2398) .. (2.9681, 1.2544) -- (2.9681, 0.7562) -- (3.3229, 0.7562) -- (3.3229, 1.4002) -- (3.4981, 1.4002) -- (3.8349, 0.7562) -- (4.2603, 0.7562) -- (3.8505, 1.4364) -- cycle(0.9628, 1.7777).. controls (0.9438, 1.7534) and (0.92, 1.7357) .. (0.8911, 1.7243).. controls (0.8623, 1.7129) and (0.83, 1.706) .. (0.7945, 1.7039).. controls (0.7588, 1.7015) and (0.7252, 1.7005) .. (0.6932, 1.7005) -- (0.4839, 1.7005) -- (0.4839, 2.0667) -- (0.716, 2.0667).. controls (0.7477, 2.0667) and (0.7805, 2.0643) .. (0.8139, 2.0598).. controls (0.8472, 2.0553) and (0.8768, 2.0466) .. (0.9025, 2.0336).. controls (0.9282, 2.0206) and (0.9496, 2.0021) .. (0.9663, 1.9778).. controls (0.9829, 1.9534) and (0.9914, 1.9209) .. (0.9914, 1.8799).. controls (0.9914, 1.8362) and (0.9819, 1.8021) .. (0.9628, 1.7777) -- cycle(2.6125, 1.3533).. controls (2.5889, 1.2904) and (2.5553, 1.2359) .. (2.5112, 1.1896).. controls (2.4672, 1.1433) and (2.4146, 1.1073) .. (2.3529, 1.0814).. controls (2.2916, 1.0554) and (2.2228, 1.0427) .. (2.1471, 1.0427).. controls (2.0712, 1.0427) and (2.0026, 1.0557) .. (1.9412, 1.0814).. controls (1.8799, 1.107) and (1.8272, 1.1433) .. (1.783, 1.1896).. controls (1.7391, 1.2359) and (1.7052, 1.2904) .. (1.6817, 1.3533).. controls (1.6581, 1.4163) and (1.6465, 1.4856) .. (1.6465, 1.5616).. controls (1.6465, 1.6359) and (1.6581, 1.705) .. (1.6817, 1.7687).. controls (1.7052, 1.8325) and (1.7388, 1.8873) .. (1.783, 1.9336).. controls (1.8269, 1.9799) and (1.8796, 2.0159) .. (1.9412, 2.0418).. controls (2.0026, 2.0675) and (2.0712, 2.0804) .. (2.1471, 2.0804).. controls (2.223, 2.0804) and (2.2916, 2.0675) .. (2.3529, 2.0418).. controls (2.4143, 2.0161) and (2.467, 1.9799) .. (2.5112, 1.9336).. controls (2.5551, 1.8873) and (2.5889, 1.8322) .. (2.6125, 1.7687).. controls (2.636, 1.705) and (2.6477, 1.6359) .. (2.6477, 1.5616).. controls (2.6477, 1.4856) and (2.636, 1.4163) .. (2.6125, 1.3533) -- cycle(3.8015, 1.7777).. controls (3.7825, 1.7534) and (3.7587, 1.7357) .. (3.7298, 1.7243).. controls (3.701, 1.7129) and (3.6687, 1.706) .. (3.6333, 1.7039).. controls (3.5975, 1.7015) and (3.5639, 1.7005) .. (3.5319, 1.7005) -- (3.3226, 1.7005) -- (3.3226, 2.0667) -- (3.5547, 2.0667).. controls (3.5864, 2.0667) and (3.6192, 2.0643) .. (3.6526, 2.0598).. controls (3.6859, 2.0553) and (3.7155, 2.0466) .. (3.7412, 2.0336).. controls (3.7669, 2.0206) and (3.7883, 2.0021) .. (3.805, 1.9778).. controls (3.8216, 1.9534) and (3.8301, 1.9209) .. (3.8301, 1.8799).. controls (3.8301, 1.8362) and (3.8206, 1.8021) .. (3.8015, 1.7777) -- cycle;
  \end{scope}
\end{tikzpicture}
}

\usepackage[affil-it]{authblk}
\usepackage{orcidlink}
\author[1%
  ]{Léo Leplat%
    \,\orcidlink{0009-0006-4178-7980}\,%
    }
\author[1%
  ]{Alan O'Callaghan%
    \,\orcidlink{0000-0003-4817-6171}\,%
    }
\author[1%
  \ensuremath\mathparagraph]{Peter Bankhead%
    \,\orcidlink{0000-0003-4851-8813}\,%
    }

\affil[1]{Institute of Genetics and Cancer, University of Edinburgh,
Edinburgh, EH4 2XU, UK%
    \,\protect\href{https://ror.org/05hygey35}{\protect\rorlogo}\,%
  }
\affil[$\mathparagraph$]{Corresponding author: %
}
\date{16 March 2026}

\begin{document}
\maketitle

\section{Summary}\label{summary}

QuPath is open-source software for bioimage analysis. As a desktop
application that is flexible and easy to install, QuPath is used by labs
worldwide to visualise and analyse large and complex images. However,
relying only on images stored only on a local file system limits
QuPath's use for larger studies.

This paper describes a new extension that enables QuPath to access
pixels and metadata from an OMERO server. This enhances the software by
allowing it to work efficiently with images stored remotely, while also
serving as a template for developers who want to connect QuPath to other
image management systems.

\section{Statement of need}\label{statement-of-need}

QuPath is a popular, open-source application for image analysis, written
in Java (\citeproc{ref-Bankhead2017}{Bankhead et al., 2017}). With over
900,000 downloads (across all releases) and more than 6,000 paper
citations to date, QuPath's support for large and complex images have
helped establish the software as a key biomedical research tool.

QuPath is routinely used to analyse whole slide images, which are common
in research and the field of digital pathology. A `small' whole slide
image might be 120,000 x 60,000 pixels in size, which equates to around
20 GB uncompressed data (assuming 8-bit, 3-channel RGB pixels). Some
images used with QuPath can be much larger, such as fluorescence
multiplexed whole slide images that often contain dozens of 16-bit or
32-bit channels. While lossy compression can somewhat reduce file sizes,
data management remains a major issue for large studies.

OMERO is a popular, open-source image management solution that enables
images to be stored on a central server and viewed through a web browser
(\citeproc{ref-Allan2012}{Allan et al., 2012}). It is installed in
institutions worldwide and is a key technology powering the
\href{https://idr.openmicroscopy.org}{Image Data Resource (IDR)}
(\citeproc{ref-Williams2017}{Williams et al., 2017}) --- a major
repository currently hosting over 400 TB of published imaging data.
OMERO also supports whole slide and multiplexed images, making it a
natural fit for many QuPath use cases.

The QuPath OMERO extension bridges the gap between both tools, making it
possible to apply QuPath analysis to images hosted in OMERO. By
efficiently accessing only the required pixels and metadata, the
extension avoids the need to download and duplicate entire datasets.

\section{State of the field}\label{state-of-the-field}

The current work builds upon lessons learned in the development of two
previous QuPath extensions that connected to OMERO.

Initially, within the QuPath development team we created the
\href{https://github.com/qupath/qupath-extension-omero-web}{QuPath Web
OMERO extension}. This lightweight extension could be added to QuPath
and used to access images through the OMERO web API. As a single jar
file without external dependencies, this was easy to install and use.
However, it had the major limitation of being able to request only
JPEG-compressed, 8-bit RGB tiles --- not raw pixel data. This made it
unsuitable for quantitative analysis and incapable of supporting most
fluorescence images, which can have different bit-depths and channels.

The \href{https://github.com/BIOP/qupath-extension-biop-omero}{BIOP
OMERO extension} started as a fork of our original extension. It was
developed by the \href{https://biop.epfl.ch/}{BioImaging and Optics
Platform (BIOP)} team to use the OMERO Ice API, rather than the web API.
This had the major advantage of raw pixel access, but also had
disadvantages: it required many additional dependencies, raw pixel
access could be slow, and it supported only OMERO servers with
authentication.

We developed the present QuPath OMERO extension from scratch to provide
a more flexible and maintainable solution. It aims to overcome previous
limitations, while adding support for recent OMERO features and
implementing best practice. Key features include the ability to:

\begin{enumerate}
\def\labelenumi{\arabic{enumi}.}
\tightlist
\item
  Browse and import images from both public and private OMERO servers.
\item
  Access pixels flexibly using different APIs.
\item
  Exchange extra information (e.g., annotated regions of interest)
  between QuPath and OMERO.
\item
  Run custom scripts within QuPath to interact with the OMERO server.
\end{enumerate}

The new extension is also the first to include unit tests. It can be
easily installed through QuPath's new extension manager, which we
developed in parallel to support downloading both the extension and its
optional dependencies.

\section{Software design}\label{software-design}

\begin{figure}
\centering
\includegraphics[keepaspectratio,alt={Screenshot showing the QuPath OMERO extension browsing the IDR.}]{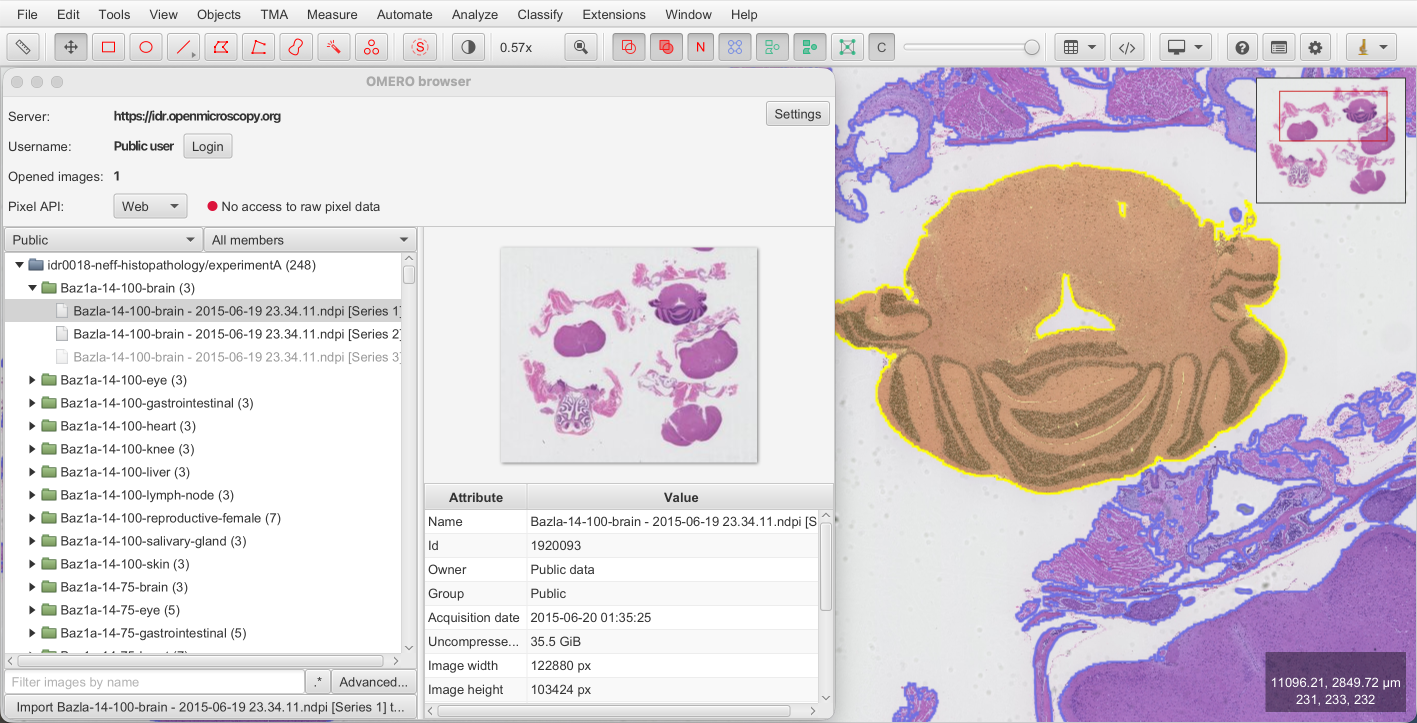}
\caption{Screenshot showing the QuPath OMERO extension browsing the
\href{https://idr.openmicroscopy.org}{IDR}.}
\end{figure}

While the need for flexible pixel access from both public and private
OMERO servers was the initial motivation for this work, we developed a
new extension (rather than another fork) to avoid the constraints of
earlier design choices and backwards compatibility. This allowed us to
better separate core logic from the user interface, improve
responsiveness, and add extensive tests. This separation is reflected in
the code being divided into two main packages: \texttt{gui} and
\texttt{core}.

The \texttt{gui} package contains the user interface controls and
interaction with the main QuPath application. It is designed to follow
the general style of QuPath and to always stay responsive by delegating
long-running tasks to background threads. When browsing an OMERO server,
the interface resembles the OMERO web client to streamline user
adoption.

The \texttt{core} package is largely independent of the user interface,
although we permitted the use of JavaFX observable properties and some
QuPath classes (e.g., to manage preferences). This was a pragmatic
decision that allowed us to write simpler code with less duplication.
The key distinction between packages is therefore that \texttt{core}
contains classes that can be used headlessly and via scripts, while
\texttt{gui} requires the interface to be instantiated.

The \texttt{core} package is also the focus of our unit testing. Because
many functions can only be tested when an active connection to an OMERO
server is established, a Docker container that hosts an OMERO server is
automatically created when unit tests are run. The extension also
provides a bash script to create this Docker container outside of the
unit tests, which can be useful for manual testing.

A central part of the \texttt{core} package is support for different
pixel APIs. Three are currently implemented:

\begin{itemize}
\tightlist
\item
  The \textbf{web} pixel API: this is enabled by default and is
  available on every OMERO server. It is fast, but suffers from the same
  limitations as the QuPath Web OMERO extension: only RGB images can be
  read, and images are JPEG-compressed.
\item
  The \textbf{Ice} pixel API: similar to the BIOP OMERO extension, this
  can read any image and access raw pixel values. It is available when
  its additional dependencies are installed and when the server requires
  authentication.
\item
  The \textbf{pixel data microservice} API: this can read any image and
  access the raw pixel values. It works for both public and private
  servers. It requires no additional installation on the client side,
  but requires that the
  \href{https://github.com/glencoesoftware/omero-ms-pixel-buffer}{OMERO
  Pixel Data Microservice} is installed on the server.
\end{itemize}

The variety of server configurations and user requirements makes
flexible support for different APIs essential. Of the three current
options, the pixel data microservice API has significant advantages from
the client's perspective, but not enough OMERO servers have installed
the necessary microservice to make it a default. Furthermore,
alternatives might become widely adopted in the future, such as the
\href{https://github.com/glencoesoftware/omero-zarr-pixel-buffer}{OMERO
Zarr Pixel Buffer}, and our extension is designed to be able to
incorporate these if required.

Instructions for the QuPath OMERO extension can be found on a dedicated
page of the
\href{https://qupath.readthedocs.io/en/stable/docs/advanced/omero.html}{QuPath
documentation}. We have also provided javadoc comments for all public
fields and methods. The javadocs are installed along with the extension
and are available via QuPath's built-in Javadoc viewer to help users
write their own scripts.

\section{Research impact statement}\label{research-impact-statement}

The QuPath OMERO extension was initially released on February 2024.
Since then, the extension has evolved through contributions, bug
reporting, and feature requests.

As of the beginning of March 2026, the QuPath OMERO extension has been
downloaded 29,727 times. This demonstrates a broad and active user
community.

The usefulness of this work is not limited to users at institutions
where images are managed using OMERO. Because OMERO is also used by
major public imaging resources, such as the IDR, the extension can help
anyone access these resources and explore the data within QuPath. If can
further serve as a template for developers to create QuPath extensions
that connect to other image management systems.

\section{AI usage disclosure}\label{ai-usage-disclosure}

No generative AI was used to write the software, documentation or paper.

\section{Acknowledgements}\label{acknowledgements}

This work was supported by the Wellcome Trust {[}223750/Z/21/Z{]}. This
project has been made possible in part by grant number 2021-237595 from
the Chan Zuckerberg Initiative DAF, an advised fund of Silicon Valley
Community Foundation.

We thank Melvin Gelbard and Rémy Dornier for contributions to earlier
QuPath OMERO extensions, and the BIOP team (especially Rémy) for
extensive testing and feedback on the current work.

For the purpose of open access, the author has applied a Creative
Commons Attribution (CC BY) licence to any Author Accepted Manuscript
version arising from this submission.

\section*{References}\label{references}
\addcontentsline{toc}{section}{References}

\protect\phantomsection\label{refs}

\end{document}